\documentclass[fleqn,usenatbib]{mnras}

\usepackage{newtxtext,newtxmath}

\usepackage[T1]{fontenc}

\DeclareRobustCommand{\VAN}[3]{#2}
\let\VANthebibliography\thebibliography
\def\thebibliography{\DeclareRobustCommand{\VAN}[3]{##3}\VANthebibliography}

\usepackage{graphicx}	
\usepackage{amsmath}	
\usepackage{orcidlink}
\usepackage{natbib}

\defcitealias{2017PotterUsingJets}{P17}
\defcitealias{2019BrombergKinkJets}{B19}
\defcitealias{Zdziarski2022}{Z22}

\title[Kink instability in outburst cycle of XRBs]{Kink instability as a particle acceleration mechanism in X-ray binaries and the connection to the outburst cycle}

\author[E. L. Elley et al.]{
Emma L. Elley$^{1,\orcidlink{0009-0002-5349-908X}}$\thanks{E-mail: emma.elley@physics.ox.ac.uk},
James H. Matthews$^{1,\orcidlink{0000-0002-3493-7737}}$,
Alex J. Cooper$^{2,3,1\orcidlink{0000-0002-4033-3139}}$,
Fraser J. Cowie$^{1,\orcidlink{0009-0009-0079-2419}}$
and Rob Fender$^{1,4,\orcidlink{0000-0002-5654-2744}}$
\\
$^{1}$Astrophysics Subdepartment, Department of Physics, University of Oxford, Keble Road, Oxford, OX13RH, UK\\
$^{2}$Trottier Space Institute at McGill, 3550 Rue University, Montreal, QC, H3A 2A7, Canada \\ 
$^{3}$Department of Physics, McGill University, 3600 Rue University, Montreal, QC, H3A 2T8, Canada\\
$^{4}$Department of Astronomy, University of Cape Town, Private Bag X3, 7701 Rondebosch, South Africa
}

\date{Accepted 2026 November 18. Received 2026 November 3; in original form 2026 September 21}

\pubyear{\the\year{}}

\begin{document}
\label{firstpage}
\pagerange{\pageref{firstpage}--\pageref{lastpage}}
\maketitle

\begin{abstract}
We consider whether the kink instability can explain particle acceleration in hard state `compact' X-ray binary jets.  We apply a 1D model to assess the expected growth rate of the instability for a variety of jet models, then scale results from simulations to estimate a dissipation rate for the magnetic energy in the jet. We then consider the effects of radiative cooling on the system. We find that hard state `compact' X-ray binary jets are likely to be unstable to the kink instability within their acceleration zones across a plausible parameter space. We further find that depending on the parameters of the model, the jets may radiate over a variety of length scales ranging from length scales attributable to a `failed' jet scenario, to those in keeping with resolved radio observations of `compact' jets. We discuss the expected evolution of this model through the outburst cycle.

\end{abstract}

\begin{keywords}
acceleration of particles -- black hole physics -- magnetic reconnection -- stars: jets -- X-rays: binaries -- radiation mechanisms: non-thermal
\end{keywords}



\section{Introduction}

Black hole (BH) X-ray binaries (XRBs) have been found to follow Q-shaped tracks in plots of X-ray intensity against hardness~\citep{2004FenderTowardsJets,2009FenderJetsX-rays}. In the hard state, the X-ray spectra are dominated by contributions from the corona, leading to a shallow photon spectral index, whereas in the soft state, the disk emission dominates. It has been argued and observationally confirmed that all BH XRBs in the hard X-ray state produce a `compact' flat spectrum jet~\citep[e.g.][]{2001FenderPowerfulStates}. Hard state `compact' jets, have been resolved in the cases of Swift J1727.8-1613 (hereafter J1727)~\citep{Wood2024SwiftBinary,2026WoodRealSwiftJ1727}, Cyg X-1 (HDE 226868, V1357 Cygni)~\citep{2001StirlingRelativisticState}, GRS 1915+105~\citep{2000DhawanAUScaleGRS, 2004RiboAsymmetricGRS}, MAXI J1836-194~\citep{2015RussellRadioMAXIJ1836}, MAXI J1820+070~\citep{2021TetarenkoMeasuringBinary}, and AT2019wey~\citep{2021YadlapalliVLBAAT2019wey}. Swift J1727, has a minimum size of $\approx5-8\times10^{-4}$ pc or $\approx1.5-2.5\times10^{15}$ cm~\citep{Wood2024SwiftBinary}, beyond which the jet drops below the limit of observability. As the sources move from the hard state to the soft state, they are often found to exhibit flaring behaviour in radio and X-rays, and to launch discrete ejecta that travel to large distances~\citep[e.g.][]{2019MillerJonesRapidlyCygni,2019RussellDiskMAXI,2020BrightAnMAXI,2023YouObservationsDisk}. In many cases flares are the result of material transitioning from an optically thick to optically thin regime~\citep{1966VanderLaanModelSources,2019FenderSynchrotronHoles, 2026CowieTowardsRegions}. The compact jet is then quenched as the source moves into the soft state~\citep[e.g.][]{2004FenderTowardsJets,2009FenderJetsX-rays,2026WoodRealSwiftJ1727}.

Discrete ejecta have been observed in many systems, travelling to large distances~\citep[][find the largest deprojected distance so far for one of these systems, in the XRB 4U 1543--47]{2026ZhangJetsHoles}. The discrete ejecta are observed to travel relativistically and to decelerate over time~\citep[e.g.][]{2022CarotenutoModellingMAXIJ1348, 2025SavardRelativisticImages}. The launch of discrete ejecta can often be associated with flaring in multiple wavelengths. Time resolved observations of flares and discrete ejecta have shown flares rising after the inferred launch of the discrete ejecta~\citep[e.g.][]{2019MillerJonesRapidlyCygni, 2019RussellDiskMAXI,2020BrightAnMAXI}, suggesting that the material involved in creating the flares may be the material making up the discrete ejecta. \cite{2026DuRadioAccretion} show that in the decaying hard state flare of GX 339-4 observed in 2010-2011, the peak of the radio emission lags behind the X-ray Compton emission peak by approximately 7 days. The radio emission is assumed to come from the hard state jet~\citep[e.g.][]{2006FenderJetsBinaries}, and so this lag suggests that the radio flare is occurring as a results of changes in the disc propagating through to the jet. Similar delays were found for MAXI J1820+070~\citep{2023YouObservationsDisk} during its 2018 outburst, which were associated with the launch of discrete ejecta~\cite{2020BrightAnMAXI}.

Most flares are thought to be caused by material transitioning from an optically thick to an optically thin regime as it propagates away from the black hole. Compared to the pure adiabatic expansion model put forward by~\cite{1966VanderLaanModelSources} to explain flaring, ongoing particle acceleration is required to explain observations from many sources, such as Cyg X-3~\citep{1973PetersonParticleOutburst} and V404 Cyg~\citep{2023FenderComprehensiveCygni}. This ongoing particle acceleration means that the number of synchrotron-emitting particles is greater at greater distances along the jet, where the lower frequency peaks are found.

Observational and theoretical constraints require particle acceleration to begin at distances close to the black hole in hard state jets($\lesssim10^{3-4}r_G$)~\citep[e.g.][]{2017GandhiElevationSystem,2020RussellRapidJ1535,2022LucchiniBhjetModel, 2023KantzasExploringJets}. Internal shocks resulting from changes in the jet velocity have been suggested as one such plausible mechanism to power particle acceleration in the hard state compact jets of X-ray binaries~\citep[e.g.][]{2000KaiserInternalMicroquasars,2003VadawaleOnGRS1915,2004TurlerQuasarGRS1915,2004FenderTowardsJets}, with flickering Lorentz factor fluctuations leading to approximately flat SEDs, as observed~\citep{2014MalzacSpectralShocks}. Interactions between jets and winds in high-mass XRBs may cause internal re-collimation shocks, leading to a further source of dissipation~\citep[e.g.][]{2010Perucho3DBinaries}. Internal shocks have further been suggested as a mechanism to power the particle acceleration in flaring sources~\citep{2018MalzacJetGX339}. Alternatively, some of the required particle acceleration may be provided by the rapid onset of the kink instability (KI) close to the jet base. The KI was discussed in relation to Poynting-flux dominated jets in~\cite{2006GianniosRoleJets} as a mechanism contributing to the bulk acceleration of the jet material past the fast magnetosonic point and has been shown to be able to effectively accelerate particles in PIC simulations of relativistic jets~\citep[e.g.][]{2020DavelaarParticleJets}. $\gamma$-rays with up to PeV energies have been detected in Cygnus X-3~\citep{2025LHAASOCygnusSource,2025LHAASOUltrahighSystems}, further motivating research into efficient particle acceleration processes in XRB sources.

In this work, we discuss the expected onset height of the KI in XRB hard state jets and use this to estimate the length scales over which energy is dissipated to non-thermal particle acceleration. In Section~\ref{sec:the mechanism}, we show that the KI is expected to grow within hard state XRB jets, estimate the dissipation properties, and investigate how these may affect the jet's propagation. In particular, in Section~\ref{sec:growth_rate_estimates}, we use a 1D jet model by~\cite{Zdziarski2022}, hereafter Z22, to consider the expected growth rates of the KI, and estimate characteristic heights and radii for the development of the instability. Once an instability has developed, the energy must be dissipated. We discuss this process in Section~\ref{sec:energy_dissipation}, considering estimates of dissipation length scales from the 1D model and from scaling the results of simulations conducted by~\cite{2019BrombergKinkJets}, hereafter B19, using our predicted characteristic heights and radii. In Section~\ref{sec:model_cycle}, we move to connecting this picture to the outburst cycle. We conclude in Section~\ref{sec:conclusion}.

This work is necessarily model-dependent. Through this approach of combining various models, we show that it is important to consider the KI in the context of XRB hard state jets. The KI is not the only mechanism likely to be at play -- as previously discussed, internal shocks have had some success in explaining many aspects of XRB hard state jets~\citep[e.g.][]{2014MalzacSpectralShocks}. The models we use discussed each make assumptions and simplifications, and in the case of the 1D model for growth rates in particular~\citepalias{Zdziarski2022}, do not capture the non-linear behaviour of the evolving system once the KI has developed. By combining these models and simulations, we can take the strengths of each and understand more about the behaviour of the system than by considering each separately. Whilst our model-dependent approach can show feasibility, to further understanding, it will be crucial for future works to simulate these systems including both the onset and development of the KI together with the radiative processes.

\section{Onset of KI in hard state XRB jets}
\label{sec:the mechanism}
An axisymmetric plasma column containing a magnetic field with a toroidal component can be subject to the KI~\citep[see e.g.][for a derivation of the stability of a plasma column]{Goedbloed_Keppens_Poedts_2019}. Considering a small radial displacement of the form
\begin{equation}
    \vec{\xi}(r,\phi,z',t) = \xi e^{i(m\phi+k'z'-\omega t)}\hat{r},
    \label{eqn:perturbation}
\end{equation}
we are interested in the $m=1$ mode of deformations, which lead to a departure from axisymmetry and the characteristic `kinked' shape. For certain combinations of magnetic and pressure field profiles, a plasma column subjected to such a deformation is unstable, such that the deformation grows quickly in amplitude and disrupts the jet structure on either a local or global scales.

We seek to evaluate growth rates for the KI in a hard state XRB jet. For this, we require an estimate of the characteristic growth timescale of the KI, $\tau'_{\rm{kink}}$, which is the characteristic comoving time for a perturbation to grow by a factor of $e$. Throughout this work, primed quantities refer to those measured in the rest frame of the fluid. \cite{Appl2000} found a comoving characteristic growth time of
\begin{equation}
    \tau'_{\rm{kink}}\approx7.5\mathcal{P'}/v'_{A,p},
    \label{eqn:characteristic_growth_time}
\end{equation}
decided by the Alfvén crossing time for the current-carrying core of the plasma column~\citep{Appl2000}, where $v'_{A,p}$ is the Alfvèn speed associated with the poloidal component. The pitch of the magnetic field, $\mathcal{P}'(r)=rB'_z(r)/B'_\phi(r)$, also plays a crucial role, with fields dominated by their toroidal component having a shorter growth timescale and as such being less stable to the KI. For now, we consider jets launched and powered by the coupling of the black hole spin to the magnetic fields crossing the horizon via the Blandford-Znajek mechanism, for which the pitch of the magnetic field is given by (\citealp{1977BlandfordElectromagneticHoles}; \citealp{2009TchekhovskoyEfficiencyMagnetosphere}; \citetalias{Zdziarski2022})
\begin{equation}
    \frac{\mathcal{P'}}{r_{\rm{H}}} \approx \frac{2\beta\Gamma}{a_*\ell},
    \label{eqn:pitch_spin}
\end{equation}
where $a_*=J/Mc$ is the dimensionless spin of a black hole with angular momentum $J$ and mass $M$, $r_H=(1+(1-a_\star^2)^{1/2})r_G$  is the reduced circumference of the event horizon for a rotating BH with $r_G=GM/c^2$, and $\ell\approx1/2$ is the ratio of the angular velocity of the field lines to the BH angular frequency. In Section~\ref{sec:growth_rate_estimates} we discuss the effects of relaxing this assumption. The Alfvèn speed is given by
\begin{equation}
    v'_{A,p}=c\sqrt{\sigma_p/(1+\sigma_p)},
    \label{eqn:Alfven_speed}
\end{equation}
and the poloidal magnetisation by
\begin{equation}
    \sigma_p=B'^2_z(r'=0)/(4\pi\rho'c^2),
    \label{eqn:magnetisation_def}
\end{equation}
where we have assumed that the jet has a small opening angle, such that the poloidal magnetic field strength is approximately equivalent to $B'_z$ across the full jet.

\subsection{Jet model}
\label{sec:jet_model}
The radius, Lorentz factor and magnetisation of a jet vary along its length, thus the growth rate of the instability will also in general vary along the jet length. To move forwards, we require a jet model covering the acceleration region. Throughout this work, where needed we assume a black hole mass of $10\rm{\,M_\odot}$, giving $r_G\approx1.5\times10^6$ cm.

To describe the shape and speed of the jet in, and shortly following, the acceleration zone, we use the 1D model of an MHD jet presented by \citetalias{Zdziarski2022}. In this model the jet transitions from a pseudo-parabolic geometry to a conical geometry as the jet reaches its maximum Lorentz factor. The four free parameters of this model are the maximum Lorentz factor $\Gamma_{\rm{max}}$, an acceleration parameter, $a_\Gamma$, controlling the height at which the maximum Lorentz factor is reached, a width parameter, $a_r$, and $q_2$, which describes how parabolic or conical the acceleration region is. A jet with $q_2 = 1$ is conical in the acceleration zone, whereas a jet with $q_2=2$ has a perfectly parabolic shape in the acceleration zone. The maximum Lorentz factor is reached at a height of $z_t=\left(\Gamma_{\rm{max}}/a_\Gamma\right)^{q_1}r_H$, where $r_H$ is the black hole event horizon radius, and $q_1=q_2/(q_2-1)$. This connection between $q_1$ and $q_2$ connects the rate of acceleration to the shape of the acceleration region and is a consequence of assuming that the product of the half opening angle of the jet, $\Theta_0$, and $\Gamma_{\rm{max}}$ is constant~\citepalias[see][for further details]{Zdziarski2022}. The radius of the jet as a function of height is given by
\begin{equation}
    r(z)= \begin{cases}
        a_r\left(\frac{z}{r_H}\right)^{\frac{1}{q_2}}r_H,& z_m\leq z\leq z_t\\
        \Theta_0[z+(q_2-1)z_t],& z>z_t
    \end{cases}\,
\end{equation}
where $z_m \equiv a_\Gamma^{-q_1}r_H$ is the height of the base of the jet, the height along the black hole rotation axis at which we consider the jet to come into existence, and 
\begin{equation}
    \Theta_0=\frac{a_r}{q_2}\left(\frac{a_\Gamma}{\Gamma_{max}}\right)^{q_1(1-\frac{1}{q_2})}.
\end{equation}
$a_r$ is a scaling factor on the width of the jet, and its value decides (together with $q_1,q_2,a_\Gamma$ and $\Gamma_{\rm{max}}$) what the half opening angle of the jet will be. A jet with a higher value of $a_r$ would be wider at any given point. We set $a_r=1$ throughout this analysis for simplicity, noting that large departures from this value result in unrealistic opening angles. $a_\Gamma$ determines the distance over which the jet reaches its final Lorentz factor. The Lorentz factor is given by
\begin{equation}
    \Gamma(z)=\begin{cases}
        a_\Gamma\left(\frac{z}{r_H}\right)^{\frac{1}{q_1}},& z_m\leq z\leq z_t\\
        \Gamma_{\rm{max}}, & z>z_t,
    \end{cases}
\end{equation}
such that the maximum Lorentz factor is reached at the height of the transition to a conical geometry. Higher values of $a_\Gamma$ result in jets which accelerate over shorter heights. The magnetisation of the jet as a function of height is given by~\citepalias{Zdziarski2022}
\begin{equation}
    \sigma(z)= \frac{\Gamma_{\rm{max}}(1+\sigma_{\rm{min}})}{\Gamma(z)}-1,
    \label{eqn:magnetisation_func_height}
\end{equation}
where $\sigma_{\rm{min}}$ is the minimum value of the magnetisation, reached at $z_t$. As a consequence of Equation~\ref{eqn:magnetisation_func_height}, $\sigma(z)$ and $\Gamma(z)$ are degenerate, once a choice of $\Gamma_{\rm{max}}$ has been made.

\subsection{Growth rate estimates}
\label{sec:growth_rate_estimates}
To calculate instability growth for jet material injected at the base, we must integrate along the length of the jet due to the dependence of the growth rate on $z$. The characteristic time varies along the jet because $\sigma_p$ and the speed of the jet change with height. Within $\tau'_{\rm{kink}}$, the jet material travels a lab frame distance of $\Gamma(z')\beta(z')c\tau'_{m=1}(z')$. Following \citetalias{Zdziarski2022}, we assume that the characteristic growth time, $\tau'_{m=1}$, is given by Equation~\ref{eqn:characteristic_growth_time}. The number of e-foldings between heights $z_1$ and $z_2$ is therefore~\citepalias{Zdziarski2022}
\begin{equation}
    \int^{z_2}_{z_1}\frac{1}{\Gamma(z')\beta(z')c\tau'_{m=1}(z')}dz'.
\end{equation}
As detailed in Section~\ref{sec:the mechanism}, the calculation of $\tau'_{m=1}(z')$ depends on the assumption of the Blandford-Znajek launching mechanism~\citep{1977BlandfordElectromagneticHoles}. The dependence of $\tau'_{\rm{kink}}$ here is purely through the linear dependence on the $\mathcal{P'}$. As such, using a different pitch angle as a result of, for example, assuming the Blandford-Payne jet launching model~\citep{1982BlandfordHydromagneticJets} instead will have a small effect on our results, likely of the same order as is introduced by the uncertainty in $a_\star$.
\begin{figure}
    \centering
    \includegraphics[width=\linewidth]{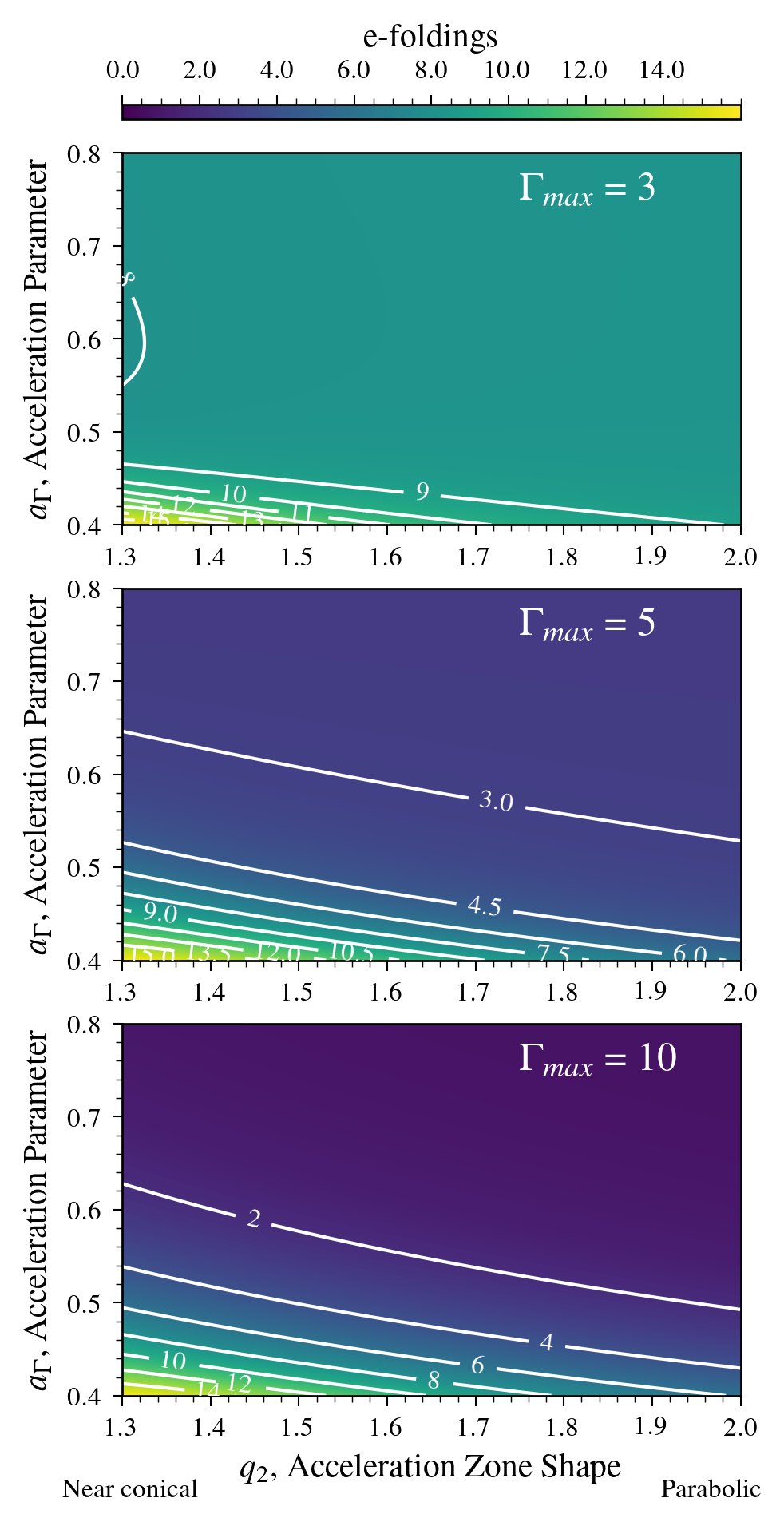}
    \caption{The contour plots show the number of times an initial perturbation grows by a factor of e by $z_{\rm{ref}} = 2000r_G$ (see panel a of Figure~\ref{fig:integration} for an illustration) for a range of acceleration parameters, $a_\Gamma$, and shapes of the acceleration region (given by $q_2$). We do this for three different values of $\Gamma_{\rm{max}}$. We begin the integration at $2z_m$. $z_m$ changes with the parameters of the model. We assume a maximally spinning black hole with a mass of $10\,\rm{M}_\odot$. A higher $q_2$ results in a jet with a narrower opening angle with an acceleration zone whose geometry is more parabolic. These more parabolic jets accelerate to $\Gamma_{\rm{max}}$ over a shorter distance, making them more stable than the conical jets with the same $a_\Gamma$. The contour labelled with 8 e-foldings in the first panel shows a different and less significant pattern (the number of e-foldings changes very slowly in this region), but nonetheless highlights a competing effect. $z_m$ changes most quickly at low $q_2$ and can have an effect on the number of e-foldings encountered, because in our scheme, a larger $z_m$ means a smaller integration region. For $\Gamma_{\rm{max}}=3$, the main takeaway is that many e-foldings are reached by $z_{\rm{ref}}$ for all of the parameter space, and all indicate growth beyond the limit of the linear theory, such that the shape of the contour showing 8 e-foldings is of little consequence to our results.}
    \label{fig:growth_rates}
\end{figure}
\begin{figure}
    \centering
    \includegraphics[width=\linewidth]{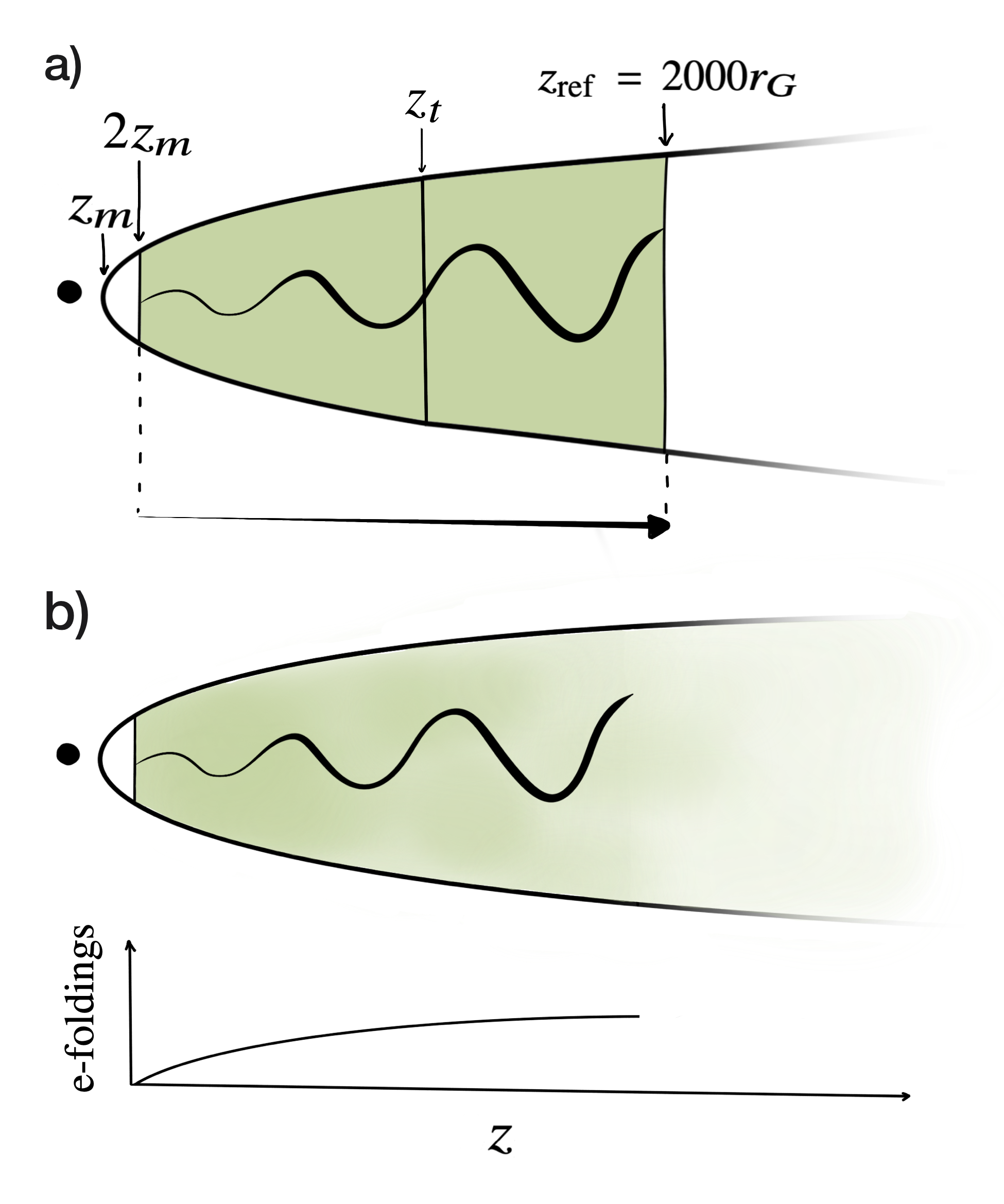}
    \caption{A perturbation to the jet material grows as the material moves along the jet axis, $z$. To produce the parameter spaces shown in Figure~\ref{fig:growth_rates}, for each given set of parameters, we integrate the number of times an initial perturbation grows by a factor of $e$ from $z=2z_m$ to $z=2000r_G$, shown here in panel a. Here we show the perturbed region as a significant fraction of the jet, but contained within the jet. Our later derivation relies on the perturbations affecting the jet material on scales of the order of the jet width, but the perturbations may or may not affect the entire jet cross section. In Figures~\ref{fig:growth_rates_by_parameter} and~\ref{fig:growth_rates_by_parameter_slow}, we integrate along the jet height, showing the cumulative growth of a perturbation from $z=2z_m$ to the height shown, illustrated here in panel b).}
    \label{fig:integration}
\end{figure}

In Figure~\ref{fig:growth_rates}, we show the number of e-foldings between $z=2z_m$ and $z = 2000r_G$, for various combinations of $q_2$, which parametrises whether the jet is conical or parabolic in the acceleration zone, and $a_\Gamma$, which controls the height at which the maximum Lorentz factor is reached. We assume a maximally spinning black hole ($a_\star=1$) in these calculations, relaxing this assumption in later sections. Panel a of Figure~\ref{fig:integration} illustrates the integration region used. Through comparison of the panels, which are made for three different values of maximum Lorentz factor, $\Gamma_{\rm{max}}$, Figure~\ref{fig:growth_rates} shows that a higher $\Gamma_{\rm{max}}$ decreases the number of e-foldings the jet experiences by $z = 2000r_G$ and therefore increases the stability of the jets. A higher value of $a_\Gamma$ implies acceleration to the maximum Lorentz factor over a shorter distance. Therefore, each individual panel of Figure~\ref{fig:growth_rates} shows that the stability of a jet decreases significantly when a jet accelerates over larger distances. A jet with a more conical acceleration zone also experiences more e-foldings than one with a more parabolic acceleration zone. Crucially, Figure~\ref{fig:growth_rates} shows that a significant number of e-foldings is expected to occur over a short distance for large areas of the available parameter space, particularly in the case of a lower maximum Lorentz factor. This suggests that the KI is an important consideration in the understanding of hard state XRB jets.

In the making of Figure~\ref{fig:growth_rates}, we have approximated the Alfvèn speed as $v'_{A,p}\approx c$. In this 1D model~\citepalias{Zdziarski2022}, as the jet accelerates, its magnetisation decreases because the magnetic energy is dissipated to kinetic energy. This gives a lower magnetisation at the end of the acceleration zone than at the very base of the jet. Therefore, in reality, the Alfvèn speed will be lower than this upper limit, such that the number of e-foldings we calculate is an upper limit. However, we expect a relatively high magnetisation to be maintained in and shortly following the acceleration region, as will be motivated by the dissipation rates for the magnetic energy density discussed in Section~\ref{sec:dissipation_simulations}, and thus adopting $v'_{A,p}\approx c$ is a reasonable approximation.

\begin{figure}
    \centering
    \includegraphics[width=\linewidth]{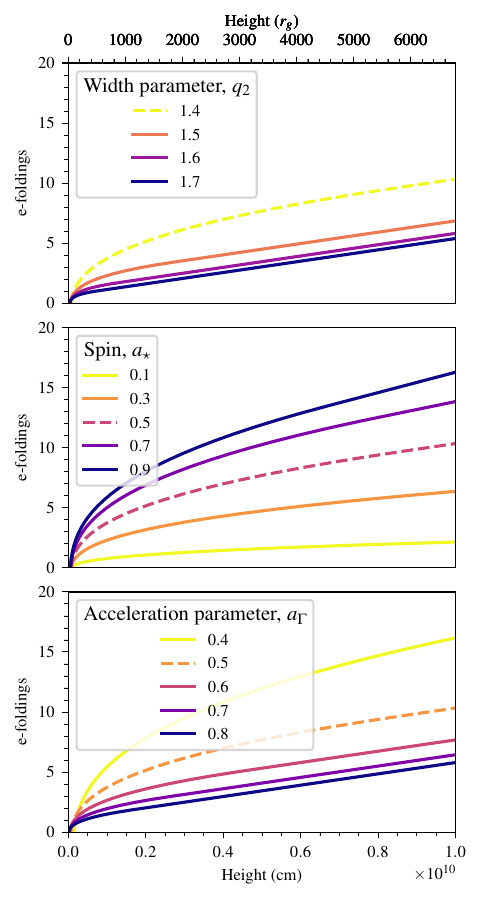}
    \caption{Growth of KI over $10^{-8}$ pc. Our fiducial model has the following parameters: $a_\star = 0.5$, $q_2 = 1.4$, $a_\Gamma = 0.5$, $\Gamma_{\rm{max}}=5$. The fiducial model is shown as a dashed line in each panel. We vary one parameter per panel as shown in the legends. Increasing $\Gamma_{\rm{max}}$ to $10$ increases the stability, however all models still reach at least $4$ e-foldings by $10^{-8}$ pc. Lowering the maximum Lorentz factor makes a given jet model less stable. We assume a black hole mass of $10\,\rm{M}_\odot$. The spin $a_*$ influences the stability predominantly through the pitch of the magnetic field. See panel b of Figure~\ref{fig:integration} for an illustration.}
    \label{fig:growth_rates_by_parameter}
\end{figure}

In Figure~\ref{fig:growth_rates_by_parameter} we take a fiducial model and vary one parameter in each panel of the figure, showing the number of e-foldings by a given height, up to a height of $1\times10^{10}$ cm. We illustrate this in panel b of Figure~\ref{fig:integration}. The most stable models have a low spin, accelerate to their maximum Lorentz factor over shorter distances and have acceleration regions which are approximately parabolic. All models shown reach a significant number of e-foldings by $1\times10^{10}$ cm, much smaller than the observed scales of hard state compact jets~\citep[e.g.][]{Wood2024SwiftBinary, 2026WoodRealSwiftJ1727}, again suggesting that the KI is an important consideration for these systems.

\begin{figure}
    \centering
    \includegraphics[width=\linewidth]{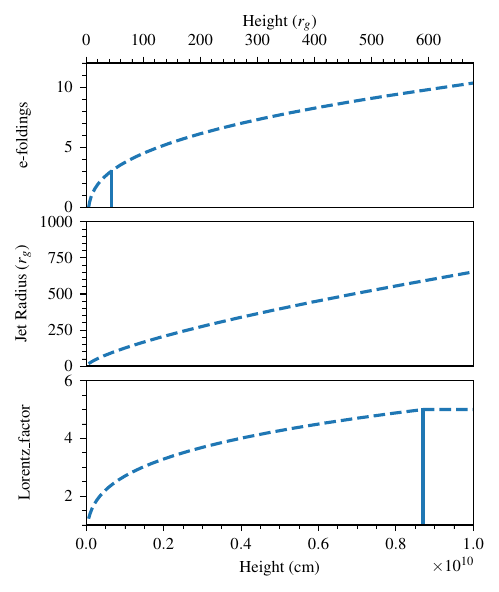}
    \caption{The e-foldings, jet radius and bulk Lorentz factor of the fiducial jet model from Figure~\ref{fig:growth_rates_by_parameter}, with the following parameters: $a_\star = 0.5$, $q_2 = 1.4$, $a_\Gamma = 0.5$, $\Gamma_{\rm{max}}=5$. The KI develops within the acceleration region of the jet, with the vertical line in the first panel showing the height at which 3 e-foldings are reached, before the height where the jet becomes conical, shown by the vertical line in the bottom panel.}
    \label{fig:fiducial_jet}
\end{figure}

\begin{figure}
    \centering
    \includegraphics[width=\linewidth]{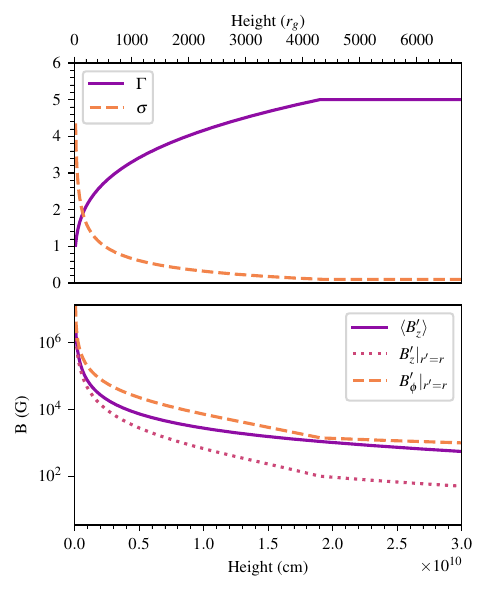}
    \caption{Properties of the jet magnetic field with height, assuming a jet power of $10^{38} \rm{erg\,s}^{-1}$ for the fiducial jet model from Figure~\ref{fig:growth_rates_by_parameter}, with the following parameters: $a_\star = 0.5$, $q_2 = 1.4$, $a_\Gamma = 0.5$, $\Gamma_{\rm{max}}=5$. $\langle B'_z\rangle$ is the poloidal magnetic field averaged across the jet radius; $B'_\phi$ is evaluated at the outer radius of the jet.}
    \label{fig:B_field_profiles}
\end{figure}

In Figure~\ref{fig:fiducial_jet}, we show the number of e-foldings, the jet radius and the bulk Lorentz factor of the fiducial jet model from Figure~\ref{fig:growth_rates_by_parameter}. In Figure~\ref{fig:B_field_profiles} we show the properties of the magnetic field for this same jet model. Here, we calculate $\sigma(z)$ using Equation~\ref{eqn:magnetisation_func_height}, assuming $\sigma_{\rm{min}}=0$ for simplicity. The top panel is illustrative of the transfer of energy from the magnetic field energy to the bulk acceleration of the jet. $\langle B'_z\rangle$ is calculated as the average of B across the jet radius as
\begin{equation}
    \langle B'_z\rangle=\frac{\Phi_B}{\pi r^2},
\end{equation}
where $\Phi_B=4\pi r_H(P_j/c)^{1/2}/a_\star l$ is the conserved poloidal magnetic flux and we assume $P_j = 10^{38} \rm{\,erg\,s}^{-1}$. Following~\citetalias{Zdziarski2022}, we calculate
\begin{equation}
    B_z'|_{r'=r}=\frac{\sigma}{1+\sigma}\langle B'_z\rangle.
\end{equation}

We use Equation~\ref{eqn:pitch_spin} to calculate
\begin{equation}
    B'_{\phi}|_{r'=r} = \frac{B_z'|_{r'=r}a_\star lr}{2r_H\beta\Gamma}.
\end{equation}
The discontinuity at $z=z_t$ in this function corresponds to the discontinuity in $\Gamma(z)$ at $z=z_t$, whereas $\langle B'_z\rangle$ is a function of the continuous function for jet radius. All values given for $B$ are proportional to $P_j^{1/2}$.


\section{Energy Dissipation}
\label{sec:energy_dissipation}
We wish to understand where in the jet particle acceleration from the KI may occur. To characterise this region we consider two quantities: $z_{\rm{diss}}$, the height at which particle acceleration begins, and $\Delta z_{\rm{diss}}$, a characteristic distance along the jet over which particles are accelerated. We begin by considering magnetic energy dissipation more broadly, describing the rate of energy transfer from magnetic energy to bulk kinetic energy and to particle acceleration, accounting for the expansion of the jet. Using this, we come to the first of two estimates for $\Delta z_{\rm{diss}}$. We then use the 1D model presented in Section~\ref{sec:jet_model} to estimate $z_{\rm{diss}}$ by looking for the height along a given jet where $3$ e-foldings of the KI have occurred, i.e. when the amplitude of a small displacement, $\xi$, has grown to $e^3\xi\approx20\xi$. The exact choice of the number of e-foldings is somewhat arbitrary, but is intended to represent a point at which the KI will have grown beyond the limits of a linear perturbation (as it is presented in Equation~\ref{eqn:perturbation}). Through this, we come to a second method of estimating $\Delta z_{\rm{diss}}$, scaling estimates from simulations to the radius of the jet given by the 1D model. We compare our results to observational constraints from measurements of the spectral break in compact XRB jets~\citep{2020RussellRapidJ1535} and to other theoretical estimates and constraints~\citep{2017PotterUsingJets,2023KantzasExploringJets}.
\subsection{Energy dissipation in 1D model}
\label{sec:dissipation_energy}
In the 1D jet model~\citepalias{Zdziarski2022} presented thus far, the acceleration is taken to be a consequence of the conversion of Poynting flux to the kinetic energy of the jet. As such, there are two reasons that the magnetic energy density of the jet changes with height: firstly, that the cross-sectional area of the jet becomes larger and the velocity higher, such that the same amount of magnetic energy is spread over a larger volume, and secondly, that the magnetic energy is being actively converted to bulk kinetic energy. The model is independent of the mechanism by which magnetic energy is transferred to bulk acceleration; however, as Figure~\ref{fig:fiducial_jet} shows, the KI is expected to grow beyond a non-linear perturbation close to the start of the acceleration zone, and as such is likely to be involved in the bulk acceleration of jet material. Such dissipation of magnetic energy has been suggested as a bulk acceleration method, particularly past the fast magnetosonic point~\citep{2006GianniosRoleJets}, where the dissipation of magnetic energy is required to allow magnetic pressure gradients to continue to accelerate the material~\citep[e.g.][]{2002DrenkhahnAccelerationDissipation}. In 2D axisymmetric models of AGN jets by~\cite{2007KomissarovMagneticJets}, the majority of the jet material has passed the fast magnetosonic surface by around $z\approx50 r_g$, which, assuming the same launching mechanism operates in BH XRB hard state jets, is well within the acceleration zone of the 1D model we use.

We consider the conservation of energy by accounting for the transfer of magnetic energy to other forms, namely kinetic energy and non-thermal energy. We do our calculations in terms of the magnetic energy density and therefore further account for the expansion of the jet:
\begin{equation}
    \frac{du_B'(z)}{dz}= \frac{du_{B,\rm{expansion}}'(z)}{dz} + \frac{du_{B\rightarrow\rm{kinetic}}'(z)}{dz} + \frac{du_{B\rightarrow\rm{nth}}'(z)}{dz}.
    \label{eqn:all_components}
\end{equation}
There is no explicit allowance in the model as presented so far for magnetic energy to be dissipated to particle acceleration. To begin with, we assume that the fraction of the energy dissipated to this form is much lower than the amount converted to kinetic energy. We return to discussing the validity of this assumption in Section~\ref{sec:radiative_cooling_effects}, where we suggest that, in order to produce the observed hard state jet morphology of, for example, Swift J1727, this assumption must hold. Below, we show that at jet heights much smaller than the transition height, the rate of energy transfer to kinetic energy with increasing height $du_{B\rightarrow\rm{kinetic}}'(z)/dz$ is proportional to the energy density of the magnetic field, $u_B$. For a non-expanding, non-accelerating fluid element, we expect a similar proportionality for the transfer of magnetic energy to non-thermal particles, albeit with a different efficiency 
\begin{equation}
    \frac{du_{B\rightarrow\rm{nth}}'(z)}{dz}\propto -u_B'.
    \label{eqn:dissipation_nth}
\end{equation}
Note that whilst we treat the rates of energy transfer from the magnetic field to other forms separately for now, they are connected through their various effects on $u'_B$ (see e.g. Equation~\ref{eqn:all_components}). More specifically, for a jet which transfers $\frac{1}{e}$ of its energy in a time $\tau'_{\rm{diss}}$, over a distance $\Gamma\beta c\tau'_{\rm{diss}}$,
\begin{equation}
    \frac{du_{B\rightarrow\rm{nth}}'(z)}{dz}= -\frac{u'_B}{\Gamma\beta c\tau'_{\rm{diss}}}.
    \label{eqn:nth_rate}
\end{equation}
As the jet approaches $\Gamma_{\rm{max}}$, the rate at which magnetic energy is converted to kinetic energy must decreases until the jet is no longer accelerating. In the ideal 1D model presented by~\citetalias{Zdziarski2022} this drops to zero at $z_t$. In a more realistic picture, dissipation of magnetic field energy may well continue, only without significant contributions to the bulk acceleration of the jet.

Combining Equations~\ref{eqn:magnetisation_func_height} and ~\ref{eqn:magnetisation_def} tells us that the magnetic energy density at a point is given by
\begin{equation}
    u_B'= \rho'c^2\left(\frac{\Gamma_{\rm{max}}(1+\sigma_{\rm{min}})}{\Gamma(z)}-1\right).
\end{equation}
To go from $\sigma_p$ to  $u_B'$ we must specify a mass density. We assume mass conservation such that $\dot{M}_j=2\pi r^2\rho'c\beta\Gamma$ is constant along $z$ allows us to write $u_B'(z)$ as a function of the initial value, $u_{B\rm{,init}}'$:
\begin{equation}
    u_B'(z) = \frac{u_{B\rm{,init}}'}{\left(\frac{\Gamma_{\rm{max}}(1+\sigma_{\rm{min}})}{\Gamma_{\rm{init}}}-1\right)}\frac{r_{\rm{init}}^2\beta_{\rm{init}}\Gamma_{\rm{init}}}{r^2(z)\beta(z)\Gamma(z)}\left(\frac{\Gamma_{\rm{max}}(1+\sigma_{\rm{min}})}{\Gamma(z)}-1\right).
\end{equation}
The derivative of this quantity with respect to $z$ has two terms. The first term relates to the spreading of the magnetic energy over a greater volume at greater heights, and the second term relates to the conversion of the magnetic energy to kinetic energy:
\begin{multline}
    \frac{du_B'(z)}{dz} = \frac{du_{B,\rm{expansion}}'(z)}{dz}+ \frac{du_{B\rightarrow\rm{kinetic}}'(z)}{dz}\\\propto \left(\frac{\Gamma_{\rm{max}}(1+\sigma_{\rm{min}})}{\Gamma(z)}-1\right)\frac{d}{dz}\left(\frac{1}{r^2(z)\beta(z)\Gamma(z)}\right)\\+ \frac{\Gamma_{\rm{max}}(1+\sigma_{\rm{min}})}{r^2(z)\beta(z)\Gamma(z)}\frac{d}{dz}\left(\frac{1}{\Gamma(z)}\right)
    \label{eqn:rate_without_nth}
\end{multline}
The first term of Equation~\ref{eqn:rate_without_nth} relating to the spreading out of the jet material, gives
\begin{multline}
    \frac{du_{B,\rm{expansion}}'(z)}{dz} \propto -\left(\frac{\Gamma_{\rm{max}}(1+\sigma_{\rm{min}})}{\Gamma}-1\right)\frac{1}{r^2z\Gamma\beta}\left(\frac{1}{q_1\beta^2}+\frac{2}{q_2}\right)
\end{multline}
for $z<z_t$ or
\begin{equation}
    \frac{du_{B,\rm{expansion}}'(z)}{dz} \propto-\frac{2\Theta}{\Gamma\beta r^3}
\end{equation}
for $z>z_t$, both proportional to $-u_B'/z$. For the second term describing the conversion of magnetic to kinetic energy:
\begin{equation}
    \frac{du_{B\rightarrow\rm{kinetic}}'(z)}{dz} \propto-\frac{\Gamma_{\rm{max}}(1+\sigma_{\rm{min}})}{q_1r^2\beta\Gamma^2z}
\end{equation}
Early in the jet, where $\Gamma(z)$ is significantly smaller than $\Gamma_{\rm{max}}$, 
\begin{equation}
    \frac{\Gamma_{\rm{max}}(1+\sigma_{\rm{min}})}{\Gamma(z)}-1\approx\frac{\Gamma_{\rm{max}}(1+\sigma_{\rm{min}})}{\Gamma(z)}.
\end{equation}
At this point, 
\begin{equation}
    \frac{du_{B\rightarrow\rm{kinetic}}'(z)}{dz}=-\frac{u_B'}{q_1}\frac{1}{\Gamma(z)z}.
    \label{eqn:dissipation_to_kinetic}
\end{equation}
$q_1$ describes how the Lorentz factor changes with height -- a lower value close to $q_1=1$ reflects a jet where $\Gamma$ is approximately proportional to $z$, whereas a higher value of $q_1$ implies quick growth of $\Gamma$ at small $z$ and slower growth at large $z$. In a jet with $q_1\approx1$, the transfer of energy to kinetic energy is at a greater rate than for greater values of $q_1$. 
The proportionality to $u_B'$ at these low values of $\Gamma$ gives this term very similar behaviour to the $\frac{du_{B\rightarrow\rm{nth}}'(z)}{dz}$ term given by Equation~\ref{eqn:nth_rate}. Conversely as $z\rightarrow z_t$, $\frac{du_{B\rightarrow\rm{kinetic}}'(z)}{dz}$ drops to low values as both $u_B'$ and the rate of change of $\Gamma(z)$ decrease. Past $z_t$, the rate must be zero as the Lorentz factor no longer changes. In Section~\ref{sec:dissipation_nth} we use this similarity in proportionality to estimate the rate at which the magnetic energy is dissipated to the acceleration of non-thermal particles.
\subsection{Dissipation of magnetic energy to non-thermal particle acceleration}
\label{sec:dissipation_nth}
We have considered how expansion of the jet material and transfer of magnetic energy to kinetic energy affect the magnetic energy density. We now revisit the conversion of magnetic energy to non-thermal particle acceleration. The three mechanisms of transfer of magnetic energy to other forms (as summarised in Equation~\ref{eqn:all_components}) will occur to different extents depending on position along the jet. For example, we saw in Section~\ref{sec:dissipation_energy} that the form of the expansion term changes at $z_t$ and that the kinetic energy term goes to $0$ at $z_t$. Similarly, the conversion of energy to particle acceleration requires the KI to have developed (or another process that can dissipate energy in this way). We do not explicitly include the transfer of energy to thermal energy. Depending on the exact bulk acceleration mechanism, there may be some transfer to and from thermal energy involved in this process. Energy will also be transferred to thermal energy whenever particles are accelerated to non-thermal energies~\citep[e.g.][]{2020DavelaarParticleJets}.
We have shown that the last two terms in Equation~\ref{eqn:all_components} have a similar form at small $z$. We assume that the transfer to non-thermal particle acceleration (and the corresponding transfer of energy to thermal energy) has a lower efficiency, and use the factor $\eta_{\rm{nth}}$ to quantify the relative fraction transferred to the energies of the particles. Combining Equations~\ref{eqn:dissipation_to_kinetic} and~\ref{eqn:nth_rate} gives
\begin{equation}
    \frac{u_B'}{\Gamma\beta c\tau_{\rm{diss}}}\approx \eta_{\rm{nth}}\frac{u_B'}{q_1}\frac{1}{\Gamma z}.
\end{equation}
This allows us to obtain an independent estimate of the dissipation timescale:
\begin{equation}
    \tau_{\rm{diss}}\approx \frac{q_1z}{\eta_{\rm{nth}}\beta c}.
\end{equation}
Taking a jet with a parabolic acceleration zone $q_2\approx2$, for $z=1-2\times10^9$ cm and $\eta_{\rm{nth}}=0.1$ this gives $\tau_{\rm{diss}}\approx1s$, assuming the jet is travelling with $\beta\approx1$. If instead, $\eta_{\rm{nth}}=0.01$, then $\tau_{\rm{diss}}\approx10s$. For material travelling at close to $c$, these dissipation times of 1 to 10 s correspond to characteristic length scales of $\Delta z_{\rm{diss}}\approx3\times10^{10}$ to $3\times10^{11}$ cm. These estimates are based on the rate of dissipation to kinetic energy at small heights (see Equation~\ref{eqn:dissipation_to_kinetic}), which must drop as $z\rightarrow z_t$. Conversely, there is evidence for ongoing particle acceleration at distances much greater than the acceleration zone~\citep[e.g.][]{1973PetersonParticleOutburst,2023FenderComprehensiveCygni}. In Section~\ref{sec:dissipation_simulations}, we find that estimates given by the comparison of the widths of jets predicted by the 1D model to simulations~\citepalias{2019BrombergKinkJets} are within an order of magnitude of the estimates given here
\subsection{Rate of energy dissipation based on scaling of simulation results}
\label{sec:dissipation_simulations}
\begin{figure*}
    \centering
    \includegraphics[width=\linewidth]{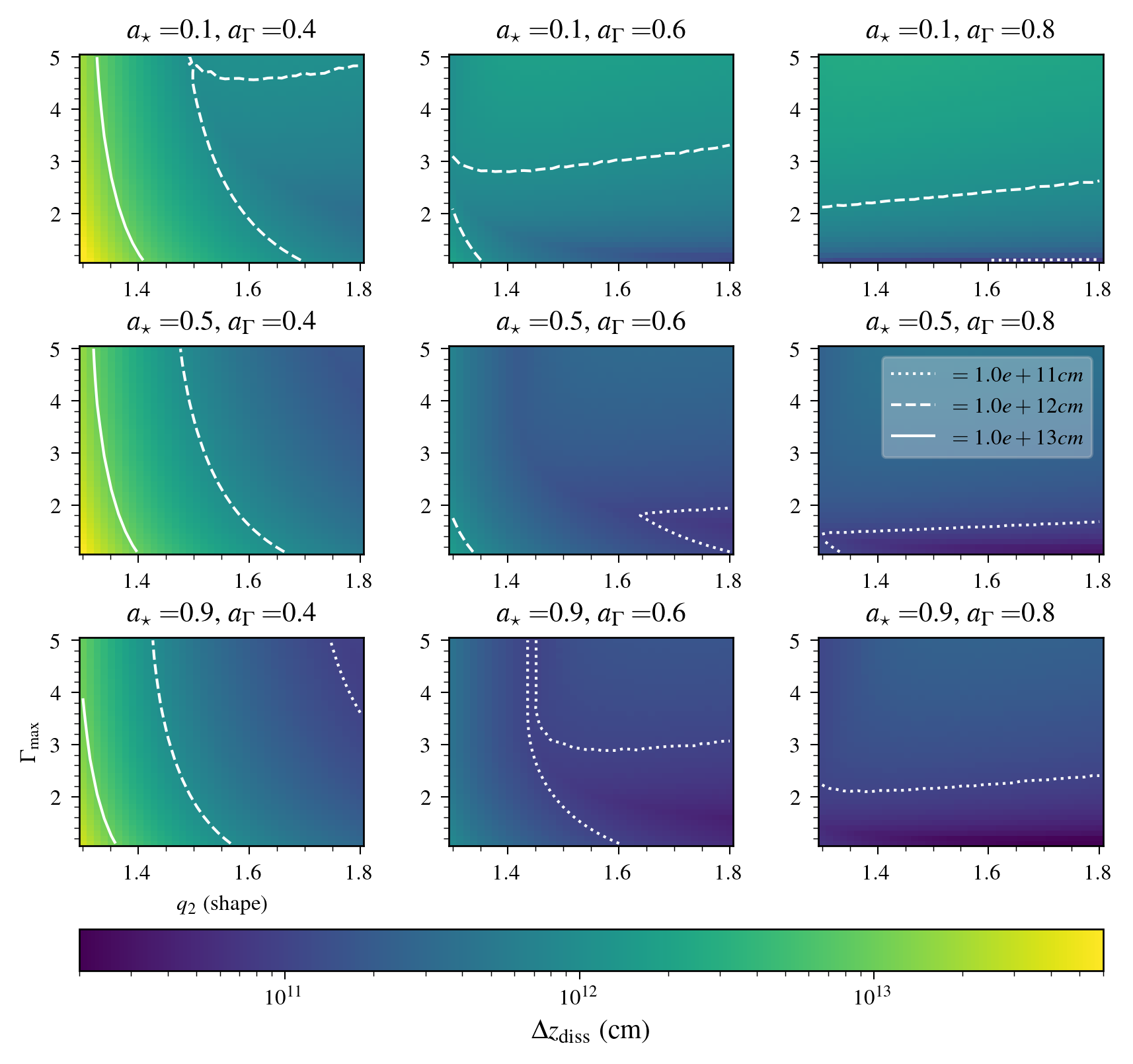}
    \caption{$\Delta z_{\rm{diss}}$ estimated via scaling a dissipation timescale suggested by simulations to the width of jets given the 1D model once the KI has developed. The dissipation length scale is calculated as $750r_j(\rm{e\,foldings} = 3)$ and we set $a_r=1$. Each individual panel shows the dissipation length scale as a function of $\Gamma_{\rm{max}}$ and $q_2$. The various panels show the dissipation length scale calculation for different pairs of $a_\star$ and $a_\Gamma$.}
    \label{fig:dissipation_lengthscales}
\end{figure*}
\begin{figure*}
    \centering
    \includegraphics[width=\linewidth]{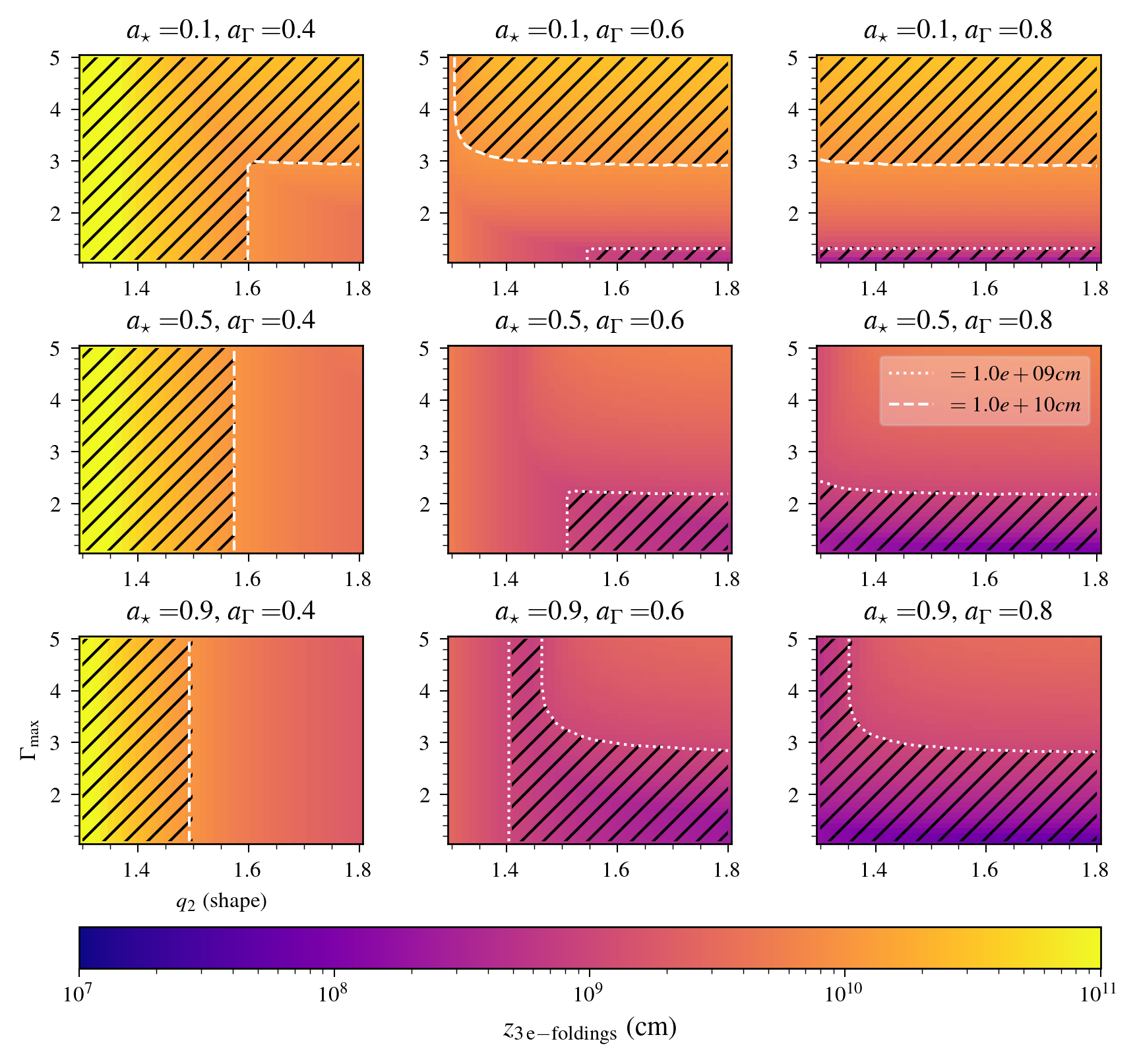}
    \caption[]{Height of 3 e-foldings with $a_r=1$. As in Figure~\ref{fig:dissipation_lengthscales}, each individual panel shows the dissipation length scale as a function of $\Gamma_{\rm{max}}$ and $q_2$. The various panels show the dissipation length scale calculation for different pairs of $a_\star$ and $a_\Gamma$. The contours and hashing refer to estimates of the height of the first acceleration zone in the compact hard state jet, taken from~\cite{2020RussellRapidJ1535}.}
    \label{fig:dissipation_heights}
\end{figure*}
Helical jet structures have been reproduced in simulations of jets covering many scales and geometries. Global simulations of GRB jets by \cite{2016BrombergRelativisticDissipation} show helical structures with many turns forming within a wider poloidal field structure. \citetalias{2019BrombergKinkJets} simulate a small section of a jet with a repeating $z$ boundary condition, finding that a helical structure develops, and in the case of a decreasing pitch profile, energy dissipation occurs rapidly as the wave number decreases instantaneously and the overall fraction of the electromagnetic energy is higher than in other cases. Thermal pressure is concentrated along the central jet axis. Approximately half of the electromagnetic energy is dissipated in the decreasing pitch case. The model we propose is similar to the simulation setup in \citetalias{2019BrombergKinkJets}, except that we imagine the jet to have some non-zero opening angle $\Theta_0$.

We use a characteristic energy dissipation timescale of $750r_j/c$ based on the simulations of \citetalias{2019BrombergKinkJets} and scale this based on jet widths of $3\times10^{8-9}$ cm, giving estimates for the dissipation timescales of $\tau_{\rm{diss}}\approx10$ to $100$ s. We note that the simulations in \citetalias{2019BrombergKinkJets} are not run to a fully relaxed state, suggesting that there will be continued dissipation at lower rates over longer timescales.

We can calculate a dissipation length scale for various combinations of the parameters of the~\citetalias{Zdziarski2022} 1D model. We do this by finding the height and radius of the 1D jet where the jet has reached 3 e-foldings. This height is used in this section as an estimate for the dissipation height, i.e. $z_{\rm{diss}}\approx z_{\rm{3\,e-foldings}}$. We then estimate a dissipation length scale as $\Delta z_{\rm{diss}}\approx750r_j(\rm{e\,foldings} = 3)$, where $r_j(\rm{e\,foldings} = 3)$ is the jet radius at the point of 3 e-foldings. Figure~\ref{fig:dissipation_lengthscales} shows the dissipation length scale calculated as a function of $q_2$ and $\Gamma_{\rm{max}}$ for various combinations of $a_\star$ and $a_\Gamma$. Furthermore Figure~\ref{fig:dissipation_lengthscales} illustrates how the scaling with $\Gamma_{\rm{max}}$ changes as a function of other jet parameters. Dissipation length scale increases with decreasing spin. This happens because the growth timescale of the KI is linearly dependent on the pitch of the magnetic field in the~\citetalias{Zdziarski2022} model, and the pitch is predicted to scale inversely with the spin of the black hole (see Equation~\ref{eqn:pitch_spin}). As $a_\Gamma$ increases, the jet accelerates to its maximum Lorentz factor over a shorter distance, meaning that they are more stable and the dissipation length scale increases. At low $a_\Gamma$, there is a strong dependence of the stability on the shape of the jet. Jets with low $a_\Gamma$ and low $q_2$ begin further from the black hole and have $z_t$ at larger heights. They are therefore generally wider than jets with higher $a_\Gamma$ and $q_2$ and therefore have larger resulting estimates for $\Delta z_{\rm{diss}}$. This dependence is particularly stark the lower the value of $a_\Gamma$. The height at which 3 e-foldings is reached can be seen in Figure~\ref{fig:dissipation_heights}, where the parameter spaces shown correspond to those in Figure~\ref{fig:dissipation_lengthscales}.

The height of the first acceleration zone in the case of hard state jets has been estimated via the spectral break. \cite{2020RussellRapidJ1535} estimate the height of the first acceleration region in MAXI J1535-571 to be approximately $10^{9-10}$ cm. We mark these two limits with contours in Figure~\ref{fig:dissipation_heights} and use hashing to exclude regions outside this range for illustrative purposes. These limits correspond to approximately 0.03 to 0.3 s for material moving at c, and so is also in broad agreement with observations of X-ray, optical and IR time lags~\citep{2017GandhiElevationSystem}. For this XRB, the observations suggest that if this 1D model can describe the hard state jet (where the jet is more compact and the corona provides a strong contribution to the overall spectrum of the XRB), then the true parameters of the hard state jet lie within the unhashed regions. \cite{2023KantzasExploringJets} use models where particle acceleration begins at heights from $z_{\rm{diss}}=100-1000\rm{\,r_g} \approx 1.5\times10^{8-9}$ cm for a $10\rm{\,M_\odot}$ black hole. Whilst the model differs from that employed here, the height of this region is similar to that found from the spectral break in MAXI J1535-571~\citep{2020RussellRapidJ1535} and from time delay measurements~\citep{2017GandhiElevationSystem}. Comparing Figures~\ref{fig:dissipation_lengthscales} and~\ref{fig:dissipation_heights} we can see that this corresponds to regions of parameter space in Figure~\ref{fig:dissipation_lengthscales} where $\Delta z\approx5\times10^{10-11}$ cm.
\subsection{Effects of radiative cooling on jet propagation}
\label{sec:radiative_cooling_effects}

Our arguments thus far have assumed that energy is conserved along the jet, and in particular that only a small fraction of the initial magnetic energy is converted to anything other than the bulk kinetic energy of the jet. Modelling of the magnetisation and radiative losses of astrophysical jets by \cite{2017PotterUsingJets}, hereafter P17, relaxes this assumption and investigates the important effects of radiative cooling of this non-thermal particle population on the jet base. \citetalias{2017PotterUsingJets} find that if the initial magnetisation is too low, the rate of energy transfer out of the jet via particle acceleration and the subsequent radiative cooling can make a significant contribution to the overall energy losses. If these losses are too high (for initial magnetisations $\sigma_B$ below $5\times10^4f_{\rm{Edd}}$, where $f_{\rm{Edd}}$ is the fractional Eddington power of the jet), then the jet may not be able to carry large fractions of its initial energy to large distances. We consider whether this may have a role in explaining the `compact' jet morphologies seen in the hard state of XRBs.

There are three potential scenarios to consider in this analysis:
\begin{enumerate}
    \item The transfer of energy to non-thermal particles and the subsequent radiative losses are insignificant compared to the energy transferred to bulk kinetic energy.
    \item The radiative losses during the acceleration zone are significant, but still allow a significant fraction of the initial energy to be transferred to the bulk kinetic energy of the jet.
    \item The radiative losses are catastrophic and result in less than $5\%$ of the initial energy going into bulk acceleration.
\end{enumerate}

The observations of resolved `compact' jets, particularly J1727~\citep{Wood2024SwiftBinary}, suggest that the third scenario is not occurring in XRB sources, as the jets are reaching at least $1.5-2.5\times10^{15}$ cm, or $\approx 10^9 r_G$ for a $10\rm{\,M_\odot}$ black hole. \citetalias{2017PotterUsingJets} find that to avoid the third scenario, the initial magnetisation of a jet needs to be $\sigma > 5\times10^4f_{\rm{Edd}}$ and that for jet powers with $f_{\rm{Edd}}>0.1$, the jet needs to remain out of equipartition until $>2\times10^4r_G$.

\citetalias{2017PotterUsingJets} further estimate a dissipation length scale of $\Delta z_{\rm{diss}}\approx10^5r_G$, equivalent to $\Delta z_{\rm{diss}}\approx3\times10^{11}$ cm for a $10M_\odot$ black hole. This estimate is dependent on multiple parameters, but results from requiring that no more than $95\%$ of the jet's initial energy is radiated before reaching equipartition. The estimate is similar to those we get both from scaling the~\citetalias{2019BrombergKinkJets} simulations and from comparing the efficiency of the dissipation to kinetic and non-thermal particle energies. Our analytic estimates in Section~\ref{sec:dissipation_nth} of $\Delta z_{\rm{diss}}\approx3\times10^{10-11}$ cm are just below the~\citetalias{2017PotterUsingJets} estimate, whereas for much of the parameter space shown in Figure~\ref{fig:dissipation_lengthscales}, estimated by scaling the simulations, we have $\Delta z_{\rm{diss}}\approx3\times10^{11-12}$ cm, which is just above the~\citetalias{2017PotterUsingJets} estimate. One approach to interpreting these results is to demand the jet has a dissipation length scale of $\Delta z_{\rm{diss}}>3\times10^{11}$ cm, in other words using the \citetalias{2017PotterUsingJets} estimate for the dissipation length scale as a minimum, and comparing this with the dissipation timescales given by our fiducial values for $\eta_{\rm{nth}}$ from Section~\ref{sec:dissipation_nth}. Doing so suggests that to explain the jets seen in XRBs, we require the efficiency of particle acceleration needs to be much lower ($\lessapprox 1\%$) than the efficiency with which energy is converted to bulk kinetic motion.

\section{Connecting the model to the outburst cycle}
\label{sec:model_cycle}

In Figure~\ref{fig:dissipation_lengthscales}, we showed that a variety of dissipation length scales may be expected depending on the jet parameters. In particular, Figure~\ref{fig:dissipation_lengthscales} shows that an increase in the maximum Lorentz factor of the jet increases the stability of the jet. This suggests that it is possible to transition from a regime where a more compact jet is expected, to a regime where the jet radiates energy over a larger distance purely by increasing the jet velocity. 

The observed compactness of a jet in VLBI observations depends heavily on observational effects. The extent of the jet observed is the extent at which the surface brightness is high enough, which is probably less than the true physical extent of the jet. An increased dissipation length scale means that energy is radiated over larger jet volume, but may mean that the surface brightness of the source decreases. Therefore, it is possible that parts of the jet dip below the surface brightness detectable by a given radio telescope. As a result the effect of an increased dissipation length scale on the observed morphology may not always directly echo the effect on the physical extent of the emitting region.

Measurements of the break frequency associated with first region of particle acceleration in the jet have been seen to move to lower frequencies during the transition from the hard state to the intermediate state in many sources~\citep[e.g.][]{2013RussellEvolvingJ1836,2013vanderHorstBroadbandJ1659,2020RussellRapidJ1535,2024EchiburuChasingModeling} and to return to high frequencies as an outburst fades and the source moves back into the hard state~\citep[e.g][]{2014RussellAccretionJ1836}. This is consistent with the first site of particle acceleration moving to larger distances from the black hole. \cite{2020RussellRapidJ1535} find that the frequency of the spectral break in MAXI J1535-571 decreased by approximately 3 orders of magnitude in a day, and showed that the rapid fading of the high-energy emission is inconsistent with radiative cooling, suggesting instead that the particle acceleration region itself is moving away from the black hole with the jet flow. High levels of variability associated with the transition out of the hard state can make estimating a spectral break difficult, however some support for this picture may be found in VLBI observations of Swift J1727. \cite{2026WoodRealSwiftJ1727} observed Swift J1727 throughout an outburst cycle, finding evidence for an extended jet morphology around the time of flaring during the hard to soft state transition. This extended morphology is on scales $\approx1000\times$ greater than our predictions for $z_{\rm{diss}}$ given in Figure~\ref{fig:dissipation_heights}. However the minimum size of the emitting region (the observed size) is not the same as the predicted $\Delta z_{\rm{diss}}$, which describes only the rate at which energy is transferred to non-thermal particles, and not the rate at which those non-thermal particles then radiate the energy. Adiabatic cooling may also greatly affect the observed morphology. Whether an increase in $z_{\rm{diss}}$ is able to influence the observed size of the jet is a complex question, both due to the observational effects mentioned previously and due to uncertainty in the distance over which the energy of the non-thermal particles is radiated.

To link our results to an outburst cycle, we consider how the transition from the hard to the soft state might affect the jet speed and magnetisation and therefore what the expected effect on $\Delta z_{\rm{diss}}$ may be. The observations discussed above hint at a picture where the jet becomes more stable at higher jet powers, with the dissipation length scale increasing and as such the bulk of the particle acceleration moving to greater distances. 

\subsection{Accretion rate during the hard to soft state transition}
\label{sec:increased_accretion_rate}
Accretion mode transitions are thought to be primarily a consequence of variations in mass accretion rate~\citep[e.g.][]{1996LasotaMechanismsTransients,1997EsinAdvectionMuscae,2004FenderTowardsJets}. The transition from a hard to soft state is often explained using a model whereby the outer edge of the central hot accretion flow (advection-dominated accretion flow, ADAF,~\citealt{1995AbramowiczThermalDisks}) moves in towards the event horizon as the accretion rate increases, because an ADAF can only be sustained at accretion rates below a critical threshold~\citep[e.g.][]{1996LasotaMechanismsTransients,1997EsinAdvectionMuscae,2014YuanHotHoles,2024JiangPhysicalBinaries}.

As the accretion rate increases, the magnetic field strength at the horizon of the black hole is expected to rise, with the effect of an increased jet power (assuming the BZ jet launching mechanism~\citealt{1977BlandfordElectromagneticHoles}) given by
\begin{equation}
    P_{\rm{jet}}= \frac{1}{24}B_h^2R_h^2ca^2,
    \label{eqn:jet_power}
\end{equation}
where $B_h$ is the magnetic field strength at the BH horizon, $a$ is the spin parameter, and $R_h$ is the radius of the BH horizon
\begin{equation}
    \frac{GM_{\rm{BH}}}{c^2}\left(1+\sqrt{1-a^2}\right).
\end{equation}
 An increase in Lorentz factor has been postulated by~\cite{2004FenderTowardsJets} and a higher accretion rate during the hard to soft state transition is supported by many observations, for example fits of mass accretion rate presented in \cite{2026DuRadioAccretion}, where the square of the central magnetic field strength $B_h^2$ is estimated to rise by a factor of $\gtrsim5$ during the decaying hard state flare. Using Equation~\ref{eqn:jet_power}, this corresponds to an increase in the jet power of $\gtrsim5\times$, which is supported by observations of the radio flux~\citep{2026DuRadioAccretion}. Furthermore, there is some observational evidence of increasing compact jet $\Gamma$ during the rising hard state~\citep{2021TetarenkoMeasuringBinary}. However, some studies have shown no significant change in bolometric luminosity during the hard to soft state transition~\citep[e.g.][]{2006CadolleBelBroadBandINTEGRAL}, suggesting that an increased accretion rate is not a universal feature of the hard-to-soft state transition.
\subsection{Effect of increased accretion rate on stability and magnetisation}
There are two major ways in which the stability of the jet can be influenced by the increased accretion rate. Firstly, the increased jet power is expected to result in an increase in the maximum Lorentz factor of the jet, with the increased kinetic power assumed to come from the conversion of a larger amount of initial Poynting flux to kinetic energy. Secondly, the increased accretion rate affects the magnetisation of the jet. We discuss these two points in the following paragraphs.

A faster jet is expected to be more stable, but as the instability sets in whilst the jet is still accelerating, we need to consider its speed as a function of the height. For a constant $a_\Gamma$, we consider the effects of increasing the accreted magnetic field from $B_{\rm{low}}$ to $B_{\rm{high}}$. The 1D model introduced in Section~\ref{sec:growth_rate_estimates}~\citepalias{Zdziarski2022} predicts that the transition height at which $\Gamma$ reaches $\Gamma_{\rm{max}}$ increases from $z_{t,\rm{low}}$ to $z_{t,\rm{high}}$. Therefore whilst the jet accelerates over a larger distances, in the jet with $B_{\rm{high}}$, the Lorentz factor up to $z_{t,\rm{low}}$ is identical to that in the $B_{\rm{low}}$ case. Past $z_{t,\rm{low}}$, stability is increased in the high magnetisation case, compared to in the low magnetisation case, because in this region, the jet material is moving more quickly. The reduced growth rate with jet speed can be seen in all curves shown in Figure~\ref{fig:growth_rates_by_parameter}: the curves turn to constant growth rates with height after the transition height is passed.

The relationship between the $\sigma(z)$ and $\Gamma(z)$ is given by~\citepalias[e.g.][]{Zdziarski2022}
\begin{equation}
    \sigma(z)=\frac{\Gamma_{\rm{max}}(1+\sigma_{\rm{min}})}{\Gamma(z)}-1,
    \label{eqn:magnetisation_z}
\end{equation}
At the jet base this implies 
\begin{equation}
    \sigma_{\rm{max}}=\Gamma_{\rm{max}}(1+\sigma_{\rm{min}})-1,
\end{equation}
giving a linear relationship between $\Gamma_{\rm{max}}$ and $\sigma_{\rm{max}}$. In other words, to reach a higher $\Gamma_{\rm{max}}$, the jet needs to have a higher initial magnetisation. We might naively expect a higher magnetisation to lead to a jet that is less stable to the KI. Equation~\ref{eqn:Alfven_speed} tells us however, that for high magnetisations (expected to be present in the base of the jet), the Alfvèn speed approaches $c$ and has only a weak dependence on the magnetisation. Together with Equation~\ref{eqn:characteristic_growth_time} this suggests that the growth rate of the KI is relatively insensitive to magnetisation in this region. 

\begin{figure}
    \centering
    \includegraphics[width=\linewidth]{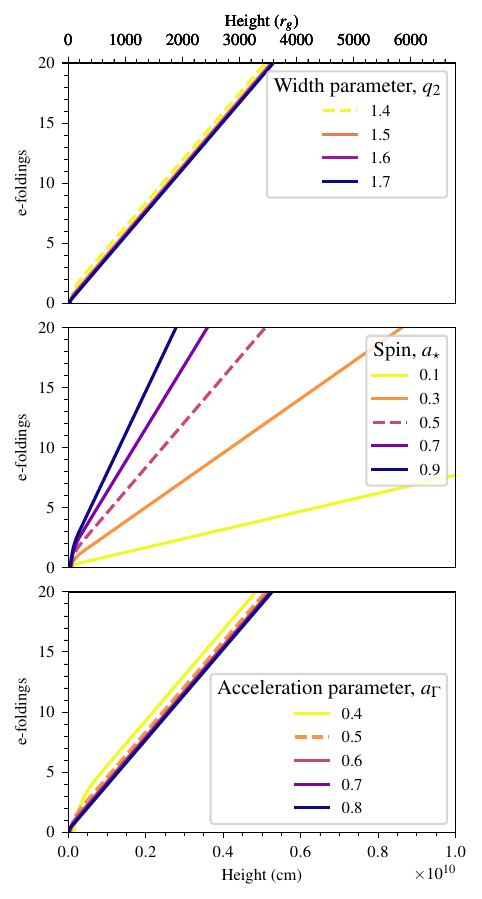}
    \caption{Same as Figure~\ref{fig:growth_rates_by_parameter} but for a final Lorentz factor of 2. Notice the change in scale on the height axis. The transition height for each curve is located at the beginning of the straight line section. These jets accelerate over a much shorter time and distance and have a smaller initial magnetisation. Whilst the magnetisation is lower, the KI grows beyond the linear limit at smaller heights, suggesting that the already lower energy in the magnetic field is transferred to other forms at a much lower height.}
    \label{fig:growth_rates_by_parameter_slow}
\end{figure}

Putting this all together, in Section~\ref{sec:growth_rate_estimates}, we showed that for most regions of parameter space for a jet with $\Gamma_{\rm{max}}=5$, the jet becomes kink unstable over a very short distance. A lower accretion rate is expected to lead to a lower final speed and a smaller transition height. We show the growth rates in the case of a final Lorentz factor of $\Gamma_{\rm{max}}=2$ in Figure~\ref{fig:growth_rates_by_parameter_slow}, finding growth beyond the linear regime at even lower heights than for $\Gamma_{\rm{max}}=5$ (see Figure~\ref{fig:growth_rates_by_parameter}), implying less stability in the $\Gamma_{\rm{max}}=2$ case. Additionally, we expect the bases of jets with lower accretion rates to have lower magnetisations.

\subsection{Crossing between the regions of parameter space}
All together, we have a situation where an increasing accretion rate is expected to trigger the system to move out of the hard state, to increase the stability of the jet to the KI and to increase the magnetisation at the jet base. We ask now what this means for the expected dissipation length scale.

As mentioned, Figure~\ref{fig:dissipation_lengthscales} shows that for many parts of parameter space, increasing the Lorentz factor by a modest amount (e.g. $\times2$) is sufficient to increase the dissipation length scale by an order of magnitude. The increased initial magnetisation expected for jets with higher final Lornetz factors offers some insight into what trends we might expect for the distances over which the energy in non-thermal particles is radiated away. A jet radiates most efficiently if it is close to equipartition ($U_B/U_e^{+-} = 1$), where efficiently means a larger fraction of the magnetic energy is radiated away, rather than remaining as magnetic energy or (later) being transferred to kinetic energy. At the jet base the jet is expected to be magnetically dominated, therefore if $U_B/U_e^{+-}$ decreases, the jet becomes closer to equipartition. Therefore for a jet with a higher initial magnetisation, not only is the jet more stable to the KI due to higher speeds, but also the dissipation of magnetic energy is less efficient, due to the jet being further from equipartition. The increased initial magnetic energy means there is more energy available to be radiated, so the overall luminosity of the jet may increase despite the decreased radiative efficiency. We might therefore expect a more extended jet to be observed during this period of high accretion rate, although as per earlier discussion, this would be subject to strong observational effects.

\subsection{Flaring}
\label{sec:flaring}
For flares, we can estimate the energy that can be released by KI and the associated magnetic reconnection for the case of a jet with increased accretion rate and compare this to observed flare energetics. The Blandford-Znajek mechanism provides an estimate for the Poynting flux as
\begin{equation}
    L_H\sim10^{45}\times\left(\frac{a}{M}\right)^2\dot{M} \rm{\, erg\,s}^{-1},
\end{equation}
where $\dot{M}$ is measured in solar masses per year. Assuming that $90\%$ of the jet energy is transferred to the bulk velocity of the jet implies an energy transferred to non-thermal particles of
\begin{equation}
    E_{\rm{flare}}\sim4\times10^{38}\rm{\,erg\,s}^{-1}\left(\frac{f_{\rm{Edd}}}{0.5}\right)\left(\frac{a_*}{0.9}\right)^2\left(\frac{ \dot{M}}{10^{-7}\rm{\,M_\odot\,yr}^{-1}}\right)\left(\frac{\Delta t}{100\rm{\,s}}\right)
\end{equation}
where $\Delta t$ is the dissipation timescale. We use a fiducial dissipation timescale of $100$ s, following our reasoning in Section~\ref{sec:dissipation_simulations}. A small amount of the initial magnetic energy will remain in the magnetic fields, but we neglect this contribution here. Our fiducial value for $\dot{M}$ is approximately the Eddington mass accretion rate for a $10\,M_\odot$ black hole.

Our resulting value of $4\times 10^{38} \rm{\,erg}$ is comparable to the mean flare energy observed in a sample of 30 candidate BH-XRBs~\citep[][Cowie et al., in prep.]{2026BaconCosmicEjecta}, implying that, energetically, the KI is a reasonable mechanism to accelerate the non-thermal particles involved in flaring. There are clear uncertainties in the percentage of the initial energy transferred to bulk acceleration, which may be larger than $90\%$, thus reducing our flare energy estimate. Conversely, the observed rise time of a flare (heavily affected by observational effects, such as observing frequency~\cite[e.g.][]{1966VanderLaanModelSources}) can be much ($\approx100\times$) longer than this dissipation timescale, such that if the KI and resulting dissipation process happens repeatedly, the expected energy can be increased.

\subsection{Discrete ejecta}
It might seem natural to expect that the discrete ejecta are caused by a jet being disrupted by a single catastrophic episode of the KI, where the rapid onset of the instability both converts magnetic energy to kinetic energy to accelerate the discrete ejecta and simultaneously leads to the catastrophic destruction of the jet structure behind the discrete ejecta. In simulations of the onset of KI in a plasma column~\citepalias[e.g.][]{2019BrombergKinkJets}, a high pressure is found towards the centre of the plasma column after the reconnection events. These simulations necessarily rely on repeating boundary conditions along the jet axis, enforcing a cylindrical geometry, whereas a physical jet is expected to expand near to its base and therefore to have a conical or parabolic geometry (as described in Section~\ref{sec:jet_model}). In a conical geometry, we might expect the increased central pressure from the development of the KI to be higher towards the base of the cone, leading to a net force on the central material along the jet direction. In the simulations~\citepalias[e.g.][]{2019BrombergKinkJets}, the large scale magnetic field is not fully destroyed by the development of the instability and may feasibly still provide some level of radial confinement to the plasma. As a consequence of the net force and confinement, we might expect some contribution to the acceleration of the discrete ejecta from the onset of the KI, which would be likely to vary in angle and speed for different ejecta. Discrete ejecta have been observed to vary in launch angle in V404 Cyg~\citep{2019MillerJonesRapidlyCygni} and Swift J1727~\citep{2026WoodRealSwiftJ1727}, suggestive of some potential underlying stochastic behaviour.

On the other hand, there are some key issues with producing discrete ejecta from catastrophic episodes of the KI. Firstly, the idea of a sudden onset of KI at higher jet powers is incompatible with our calculations as we have shown that we expect more powerful jets to be more stable to the onset of the KI, or equivalently that we expect the KI in more powerful jets to grow over larger distances. Rather our calculations suggest that almost all areas of parameter space are unstable to KI and that it should be even more prevalent in the hard state compact jets than in more powerful jets. Conversely, the increase in stability may bring the onset of the KI into a more stochastic regime, more consistent with flaring behaviour. Secondly, whilst simulations sometimes show the destruction of the full jet structure due to the onset of the KI, this is by no means inevitable. It is possible also for the outer regions to remain stable whilst KI grows in the innermost region of a jet~\citep{2016BrombergRelativisticDissipation} or for KI to grow on smaller scales throughout a region of a jet. Previous simulations of AGN jets~\citep{2026ElleyImpactJets} have shown that the KI can form a discrete boundary between regions of a jet (as required to form discrete ejecta), but this relies on having a particular field geometry and balance between magnetic and thermal pressures that is likely to be very different towards the base of an XRB jet. Furthermore, even if such a discrete boundary is formed, the jet power still needs to switch off to prevent the jet reforming behind the boundary.

Instead, it makes more sense for the the launch of discrete ejecta to be caused by a very brief and intense increase in jet power. Further simulation work would be needed to understand whether this is a feasible method of producing a discrete ejecta. An immediate challenge to contend with is the incredibly high power requirements for discrete ejecta to propagate as far as they do~\citep[e.g.][]{2024CarotenutoConstrainingModels,2025SavardRelativisticImages}. Further to this, any such simulation or group of simulations would need to cover a large range of scales, from the acceleration zone, right out to the propagating ejecta. The inclusion of magnetic fields may protect the ejecta from the development of hydrodynamic instabilities, allowing it to propagate further, as suggested by~\cite{2025SavardRelativisticImages}.

\section{Conclusions}
\label{sec:conclusion}
We have discussed a model in which particle acceleration in the hard state jet involves the onset of KI in the jet base. We showed that significant growth of the KI is expected to develop near to the base of X-ray binary jets in Section~\ref{sec:growth_rate_estimates} using a 1D jet model from~\citetalias{Zdziarski2022}. We estimated the dissipation rate by scaling simulations of the onset of KI from~\citetalias{2019BrombergKinkJets} in Section~\ref{sec:dissipation_simulations}, showing that the characteristic dissipation height is expected to increase with jet power. In Section~\ref{sec:dissipation_nth} we consider the dissipation of energy to non-thermal particle acceleration compared to the magnetic energy converted to bulk acceleration, finding that for jets with small fractions of energy going to non-thermal particles, the dissipation times are similar to those predicted by the simulation scaling results of Section~\ref{sec:dissipation_simulations}. By considering a model for radiative cooling in jets, we show that for large areas of parameter space, hard state XRB jets are expected to dissipate magnetic energy to non-thermal particles over an extended region, and that this is required for them to radiate over larger distances and have extended morphologies, such as those seen in observations of resolved, `compact' jets~\citep[e.g.][]{Wood2024SwiftBinary, 2026WoodRealSwiftJ1727}. In Section~\ref{sec:model_cycle}, we have argued that the increase in accretion rate expected at the hard to soft state transition is likely to increase both the initial magnetisation and final Lorentz factor of the jet, leading to a more extended compact jet morphology.

We summarise our main results as follows:

\begin{itemize}
    \item The KI is expected to grow beyond linear theory within the bulk acceleration zone of hard state XRB jets. The KI is therefore expected to lead to the dissipation of magnetic energy to non-thermal particle acceleration on scales comparable to the size of the bulk acceleration zone and beyond.
    \item Increasing accretion rates lead to jets which dissipate energy via the kink instability over greater distances. As a result, we expect more elongated morphologies as the jet transitions from the hard to soft state in sensitive and resolved radio observations.
    \item For low accretion rates, the KI may lead to a `failed' jet where a low initial magnetisation and quick instability growth rate lead to efficient dissipation of jet energy to radiative power on very small scales.
    \item The KI can provide the required amount of energy for flaring within the framework we present.
    \item Modest changes in Lorentz factor can lead to dramatic changes in the stability of the hard state jet. 
\end{itemize}

Our results show that the KI is likely to play an important role in explaining XRB phenomenology, in agreement with previous work~\citep[e.g.][]{2006GianniosRoleJets}. This does not preclude the involvement of other mechanisms, for example internal shocks~\citep[e.g.][]{2014MalzacSpectralShocks}. The true picture is likely to be more complex, involving both mechanisms and likely requiring further simulation work to understand, incorporating jet launching physics, together with modelling of particle acceleration and observational effects.
\section*{Acknowledgements}
JHM and ELE acknowledge funding from a Royal Society University Research Fellowship (URF\textbackslash R1\textbackslash221062). AJC acknowledges support from the Oxford Hintze Centre for Astrophysical Surveys which is funded through generous support from the Hintze Family Charitable Foundation. FJC was supported by STFC grant ST/Y509474/1. RF acknowledges support from UKRI, ERC Synergy Grant ‘Blackholistic’ (grant agreement no.
101071643) and The Hintze Family Charitable Foundation. We are grateful for the use of the following software packages: matplotlib~\citep{Hunter2007Matplotlib:Environment}, pandas~\citep{McKinney2010DataPython,Thepandasdevelopmentteam2024Pandas-dev/pandas:Pandas}, scipy~\citep{Virtanen2020SciPyPython}. 

\section*{Data Availability}

There are no new data associated with this work.



\bibliographystyle{mnras}
\bibliography{KIreferences} 

\appendix

\bsp	
\label{lastpage}
\end{document}